\documentclass{article}
\usepackage{spconf,amsmath,graphicx}
\usepackage[export]{adjustbox}
\usepackage{listings}
\usepackage{xcolor}
\usepackage{tikz}
\usepackage{threeparttable}
\usetikzlibrary{positioning,arrows.meta}
\lstdefinestyle{promptbox}{
  basicstyle=\ttfamily\footnotesize,
  columns=fullflexible,
  breaklines=true,
  breakatwhitespace=false,
  frame=tb,
  framerule=0.5pt,
  numbers=none,
  linewidth=\columnwidth,
  xleftmargin=0pt,
  xrightmargin=0pt,
  framesep=3pt,
}
\usepackage{xurl}

\title{The TMU System for the XACLE Challenge: Training Large Audio Language Models with CLAP Pseudo-Labels}
\name{Ayuto Tsutsumi$^{\star}$, Kohei Tanaka$^{\star}$, Sayaka Shiota\thanks{$^{\star}$These authors contributed equally to this work.}}
\address{Tokyo Metropolitan University, Tokyo, Japan}

\begin{document}
\ninept
\maketitle
\begin{abstract}
In this paper, we propose a submission to the x-to-audio alignment (XACLE) challenge. The goal is to predict semantic alignment of a given general audio and text pair. The proposed system is based on a large audio language model (LALM) architecture.
We employ a three-stage training pipeline: automated audio captioning pretraining, pretraining with CLAP pseudo-labels, and fine-tuning on the XACLE dataset.
Our experiments show that pretraining with CLAP pseudo-labels is the primary performance driver.
On the XACLE test set, our system reaches an SRCC of 0.632, significantly outperforming the baseline system (0.334) and securing third place in the challenge team ranking. Code and models can be found at \url{https://github.com/shiotalab-tmu/tmu-xacle2026}.
\end{abstract}


\section{Introduction}
\label{sec:introduction}
The goal of the XACLE challenge~\cite{XACLE2026} is to build a model that automatically predicts alignment scores between audio and text for text-to-audio evaluation. The objective is to achieve evaluations that correlate highly with human subjective assessments.

Motivated by the success of NLP reward models~\cite{ziegler2019fine, ouyang2022training} and the recent achievements of LALMs in various audio understanding tasks~\cite{chu2024qwen2, ghosh2025audio, team2023gemini}, we regard the XACLE challenge as reward modeling for audio-text pairs.
In this work, we construct an LALM-based model that regresses alignment scores from audio-text pairs.
Our LALM combines an audio encoder with a large language model (LLM) and is fine-tuned on the small-scale XACLE dataset.
% This is motivated by XACLE's limited data size and the fact that Whisper~\cite{whisper} encoders commonly used in LALMs~\cite{chu2024qwen2, ghosh2025audio, ding2025kimi} are not well-suited for the XACLE domain.
Specifically, we employ BEATs~\cite{pmlr-v202-chen23ag}, which has shown strong performance on environmental sound classification, as the audio encoder and Qwen2.5-0.5B~\cite{qwen2.5} as the LLM, connected via an audio projection network.
To train this LALM effectively, a three-stage training pipeline is employed. It consists of automated audio captioning (AAC)~\cite{aac} pretraining, weakly supervised learning with pseudo-labels from CLAP scores~\cite{elizalde2023clap}, and fine-tuning on the XACLE Challenge 2026 dataset training set.

Our primary contribution is demonstrating that pretraining with CLAP pseudo-labels enables effective training of LALMs for audio-text alignment scoring. Notably, the trained LALM surpasses the teacher CLAP model, demonstrating the architectural advantages of LALMs for this task.

\section{System}
\label{sec:method}

\subsection{Model Architecture}
\label{ssec:model}
\begin{figure}[tb]
  \includegraphics[bb=0 0 380 272, clip, width=\linewidth]{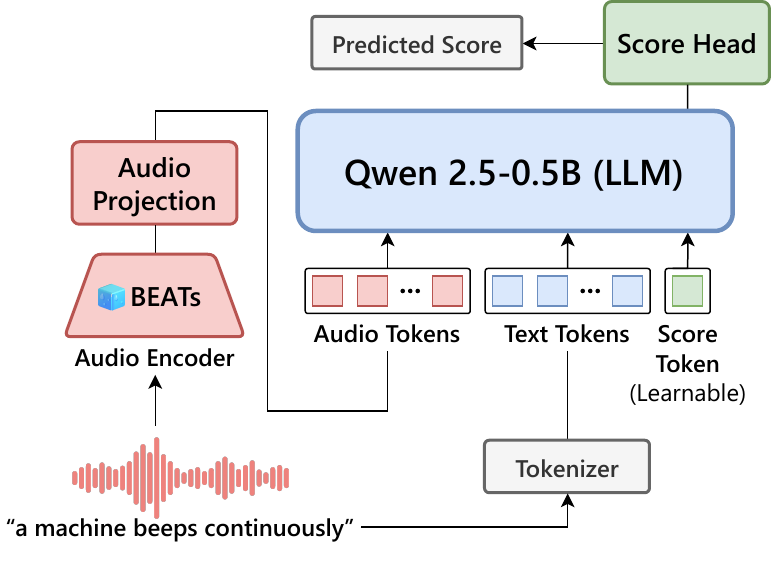}
\caption{Overview of the proposed system. Frozen BEATs features are projected into the LLM, and the alignment score is regressed from the Score Token.}
\label{fig:model}
\end{figure}

As shown in Fig.~\ref{fig:model}, the proposed model consists of an audio encoder, an audio projection, an LLM, and a score head.
Table~\ref{tab:model_components} presents the component specifications and model configurations.

\begin{table}[tb]
\centering
\caption{Model architecture and component specifications}
\label{tab:model_components}

\vspace{0.2em}

\begin{tabular}{llr}
\hline
\textbf{Component} & \textbf{Configuration} & \textbf{Params} \\
\hline
Audio Encoder & BEATs-iter3+ (768-dim, frozen) & 90M \\
Audio Projection & 3 Layers MLP, 100 tokens & 10M \\
LLM & Qwen2.5-0.5B (896-dim) & 494M \\
Score Head & Linear (896→1) & 1K \\
\hline
Total & & 594M \\
\hline
\end{tabular}
\end{table}

\noindent\textbf{Audio Encoder:}
BEATs-iter3+~\cite{pmlr-v202-chen23ag} (frozen) converts a 10\,s waveform into a sequence of 768-dimensional token embeddings at 50 tokens/s (500 tokens total).

\noindent\textbf{Audio Projection:}
The BEATs output is downsampled to 100 tokens via temporal average pooling, then transformed by a 3-layer SwiGLU MLP~\cite{shazeer2020glu} to match the LLM's input space.
Preliminary experiments comparing Q-Former~\cite{10.5555/3618408.3619222} and MLP projections (1-3 layers) showed that this configuration achieved the best performance.

\noindent\textbf{LLM:}
For the LLM, Qwen2.5-0.5B-Instruct is used.
To enable the LLM to recognize audio tokens, three special tokens are added to the vocabulary, namely \texttt{<|AUDIO\_START|>} and \texttt{<|AUDIO\_END|>} to mark the beginning and end of audio tokens, and \texttt{<|SCORE|>} for score prediction.
The input to the LLM follows the format:
\vspace{-0.8\baselineskip}
\begin{center}
\ttfamily\footnotesize
\texttt{Text~Tokens... <|AUDIO\_START|> Audio~Tokens... <|AUDIO\_END|> <|SCORE|>}
\normalfont
\end{center}
\vspace{-0.8\baselineskip}

\noindent\textbf{Score Head:}
To extract task-specific embeddings from the LLM, we use an explicit \texttt{<|SCORE|>} task token at the sequence end.
The output at the \texttt{<|SCORE|>} position is fed into a linear layer to obtain the score.
While this design is similar to the CLS token in BERT~\cite{BERT}, it is placed at the end due to causal attention~\cite{wang2024improving}.
Preliminary experiments showed this approach outperformed using the last token's hidden state~\cite{E5-V}.

\subsection{Training Pipeline}

\label{ssec:training}
\begin{figure}[t]
  \includegraphics[bb=0 0 2200 1200, clip, width=\linewidth]{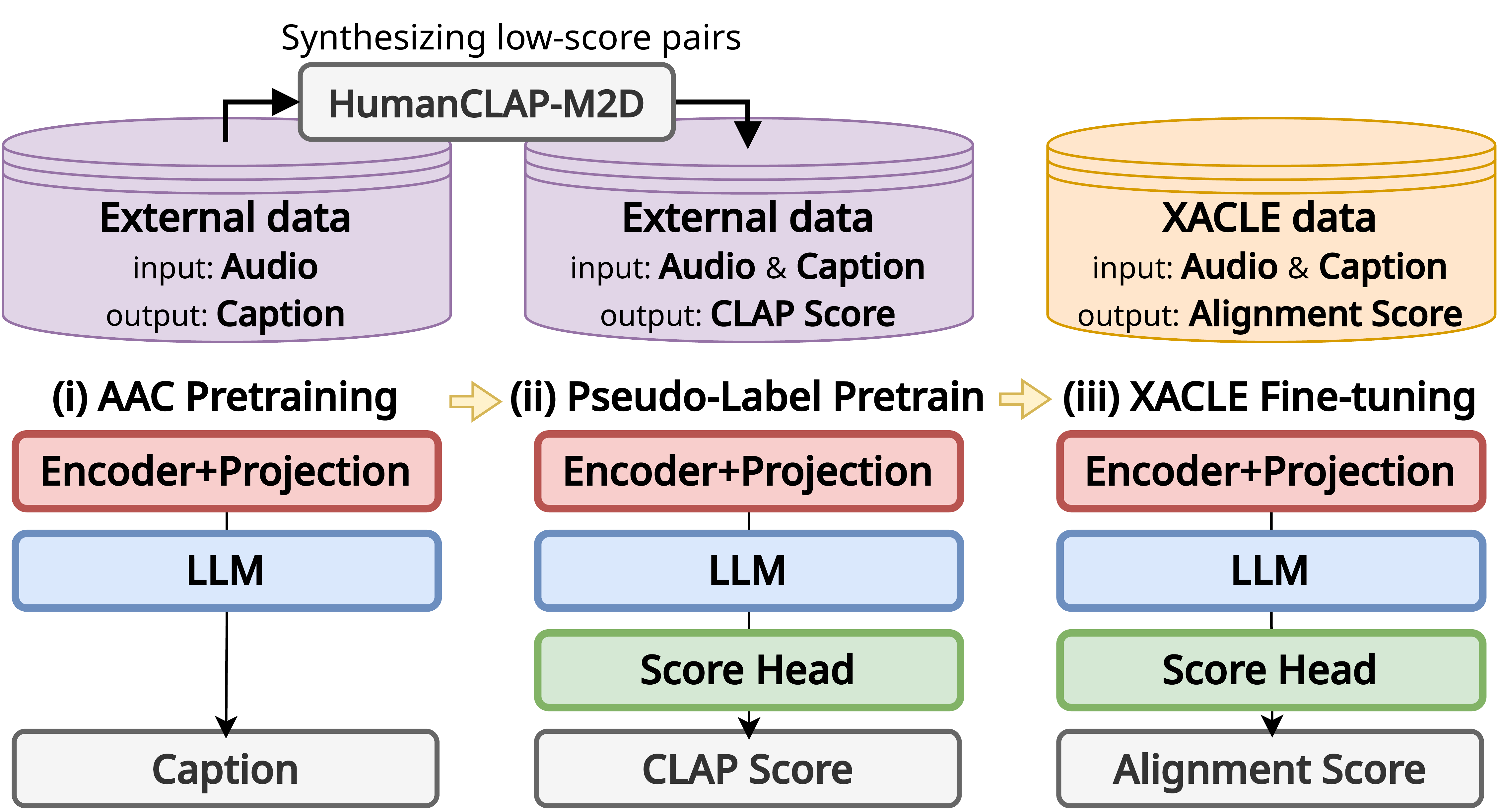}
\caption{Three-stage training pipeline. We first pretrain the projection and LLM for AAC, then introduce a score head and pretrain with CLAP pseudo-labels, and finally fine-tune on XACLE human alignment scores.}
\label{fig:train}
\end{figure}

As depicted in Fig.~\ref{fig:train}, we train the proposed model in three stages.

\noindent\textbf{Stage 1: AAC Pretraining.}
The LLM and audio projection are pretrained on the AAC task using 273K samples from AudioCaps~\cite{kim-etal-2019-audiocaps} and a subset of AudioSetCaps~\cite{bai2024audiosetcapsnipsws}, sourced from VGGSound~\cite{VGGSound}.

\noindent\textbf{Stage 2: Pretraining with CLAP Pseudo-Labels.}
We add a score head to the pretrained LLM in Stage 1 and perform weakly supervised transfer learning for the CLAP score prediction task.
Since AAC datasets contain only matched pairs, they would yield higher alignment scores than the XACLE distribution. We augment the data by synthesizing low-score pairs through negative sampling.
For audio-text pairs from the external data used in Stage~1, we replace either the audio or text with content from different samples, expanding the dataset to approximately 1,064K training samples.
We adopt HumanCLAP-M2D, which is M2D-CLAP~\cite{10502167} fine-tuned using the HumanCLAP~\cite{humanclap} method on the XACLE Challenge 2026 dataset training set, for generating pseudo-labels.
Since rank relationships are more important than absolute score values in this task, we adopt ListNet~\cite{listnet}, a loss function that directly optimizes score ordering, to ensure training is consistent with the evaluation metric SRCC.

\noindent\textbf{Stage 3: XACLE Fine-tuning.}
The model is fine-tuned on the XACLE Challenge 2026 dataset training set (7.5K samples). Due to the limited amount of training data, SpecAugment~\cite{park19e_interspeech} is applied as data augmentation to improve generalization performance and stabilize training.
ListNet is also used as the loss function.

\section{Experiments and Results}
\label{sec:results}
All models are trained using the AdamW optimizer with batch size 16.
Learning rates are 1e-5 for Stages 1 and 2 and 6.2e-6 for Stage 3.
We train for 3, 20, and 150 epochs in Stages 1, 2, and 3, respectively.
For SpecAugment in Stage 3, frequency masking of 15 and time masking of 30 are applied.
For the test set submission, we selected Stage 3 (full pipeline), No AAC Pretraining (trained from Stage 2, skipping Stage 1), and an ensemble combining these two through rank averaging: predicted scores are converted to ranks, averaged, and linearly scaled to the final score range.

Table~\ref{tab:results} shows the SRCC scores on the validation and test sets. On the validation set, each stage progressively improves performance: from Stage 1 (AAC Pretraining) to Stage 2 (Pseudo-Label Pretraining), the SRCC increases to 0.598, and adding Stage 3 (XACLE Fine-tuning) further improves the SRCC to 0.674. This demonstrates that each pretraining stage contributes to better alignment with human quality judgments. Notably, Stage 2's score of 0.598 closely matches the teacher model HumanCLAP-M2D's score of 0.602, indicating that weak supervision with pseudo-labels effectively learns alignment comparable to the teacher model.
Both the validation and test sets show a similar trend: the SRCC for No AAC Pretraining is nearly identical to that of the full pipeline (0.626 vs. 0.625). This suggests that Stage 1 may not be essential for the main task. Finally, on the test set, rank-averaging the full pipeline and the No AAC Pretraining variant yields the best test SRCC of 0.632, indicating complementary errors between the two models. We attribute this to task mismatch: AAC does not transfer well to discriminative scoring. This is supported by our observation that score-trained models lose their ability to generate captions. These results suggest that direct alignment-focused pretraining is more critical than generative pretraining for the task.

\begin{table}[t]
\centering
\begin{threeparttable}
\caption{SRCCs on XACLE test and validation sets}
\label{tab:results}
\begin{tabular}{lcc}
\hline
\textbf{Configuration} & \textbf{Val} & \textbf{Test} \\
\hline
Official Baseline & 0.384 & 0.334  \\
Stage 1: AAC Pretraining\tnote{$\dagger$} & 0.352 & - \\
Stage 1 \& 2: Pseudo-Label Pretraining & 0.598 & -  \\
Stage 1, 2 \& 3: XACLE Fine-tuning & \textbf{0.674} & 0.625 \\
\hline
No Pretraining (Stage 3 only) & 0.574 & - \\
HumanCLAP-M2D\tnote{$\ddagger$} & 0.602 & - \\
No AAC Pretraining (Stage 2 \& 3 only) & 0.669 & \textbf{0.626} \\
\hline
Ensemble & \textbf{0.678} & \textbf{0.632} \\
\hline
\end{tabular}
\begin{tablenotes}
\footnotesize
\item[$\dagger$] The cosine similarity of text embeddings between generated captions and ground-truth captions is evaluated as the score.
\item[$\ddagger$] Teacher model evaluated directly on XACLE; used for generating pseudo-labels in Stage 2.
\end{tablenotes}
\end{threeparttable}
\end{table}

\section{Conclusion}
\label{sec:conclusion}
This paper proposes a LALM-based system to predict semantic alignment between audio-text pairs for the XACLE challenge.
The proposed system leverages BEATs as the audio encoder and Qwen2.5 as the LLM, trained through a three-stage pipeline that utilizes extensive external data.
Our analysis reveals important insights: pretraining with CLAP pseudo-labels proves highly effective, with the LALM achieving an SRCC of 0.674 on validation compared to the teacher model's 0.602. This suggests that LALM-based approaches offer architectural advantages over CLAP-based methods for alignment scoring.
The final ensemble system achieves an SRCC of 0.632 on the test set, significantly outperforming the baseline and obtaining third place in the challenge, validating the effectiveness of the proposed approach.

% \vfill\pagebreak

\bibliographystyle{IEEEbib}
\bibliography{strings,refs}

@INPROCEEDINGS{humanclap,
  author={Takano, Taisei and others},
  booktitle={APSIPA ASC}, 
  title={Human-CLAP: Human-perception-based Contrastive Language-audio Pretraining}, 
  year={2025},
  volume={},
  number={},
  pages={131-136},
  doi={10.1109/APSIPAASC65261.2025.11249121}}

@inproceedings{listnet,
author = {Cao, Zhe and others},
title = {Learning to rank: from pairwise approach to listwise approach},
year = {2007},
isbn = {9781595937933},
address = {New York, NY, USA},
url = {https://doi.org/10.1145/1273496.1273513},
doi = {10.1145/1273496.1273513},
booktitle = {ICML},
pages = {129–136},
numpages = {8},
location = {Corvalis, Oregon, USA},
}

@inproceedings{
  ghosh2025audio,
  title={Audio Flamingo 3: Advancing Audio Intelligence with Fully Open Large Audio Language Models},
  author={Sreyan Ghosh and others},
  booktitle={NeurIPS},
  year={2025},
  url={https://openreview.net/forum?id=FjByDpDVIO}
}

@article{chu2024qwen2,
  title={Qwen2-audio technical report},
  author={Chu, Yunfei and others},
  journal={arXiv preprint arXiv:2407.10759},
  year={2024}
}

@InProceedings{pmlr-v202-chen23ag,
  title = 	 {{BEAT}s: Audio Pre-Training with Acoustic Tokenizers},
  author =       {Chen, Sanyuan and others},
  booktitle = 	 {ICML},
  pages = 	 {5178--5193},
  year = 	 {2023},
  url = 	 {https://proceedings.mlr.press/v202/chen23ag.html},
}

@inproceedings{10.5555/3618408.3619222,
author = {Li, Junnan and others},
title = {BLIP-2: bootstrapping language-image pre-training with frozen image encoders and large language models},
year = {2023},
booktitle = {ICML},
articleno = {814},
numpages = {13},
location = {Honolulu, Hawaii, USA},
}

@inproceedings{kim-etal-2019-audiocaps,
    title = "{A}udio{C}aps: Generating Captions for Audios in The Wild",
    author = "Kim, Chris Dongjoo and others",
    booktitle = "NAACL-HLT",
    year = "2019",
    address = "Minneapolis, Minnesota",
    url = "https://aclanthology.org/N19-1011/",
    doi = "10.18653/v1/N19-1011",
    pages = "119--132",
}

@inproceedings{
bai2024audiosetcapsnipsws,
title={AudioSetCaps: Enriched Audio Captioning Dataset Generation Using Large Audio Language Models},
author={Bai, Jisheng and others},
booktitle={Audio Imagination: NeurIPS 2024 Workshop AI-Driven Speech, Music, and Sound Generation},
year={2024},
url={https://openreview.net/forum?id=uez4PMZwzP}
}

@ARTICLE{10502167,
  author={Niizumi, Daisuke and others},
  journal={IEEE/ACM Trans. Audio, Speech, Lang. Process.},
  title={Masked Modeling Duo: Towards a Universal Audio Pre-Training Framework},
  year={2024},
  number={},
  pages={2391-2406},
  doi={10.1109/TASLP.2024.3389636}}

@inproceedings{park19e_interspeech,
  title     = {SpecAugment: A Simple Data Augmentation Method for Automatic Speech Recognition},
  author    = {Daniel S. Park and others},
  year      = {2019},
  booktitle = {Interspeech},
  pages     = {2613--2617},
  doi       = {10.21437/Interspeech.2019-2680},
  issn      = {2958-1796},
}

@article{ouyang2022training,
  title={Training language models to follow instructions with human feedback},
  author={Ouyang, Long and others},
  journal={NeurIPS},
  pages={27730--27744},
  year={2022}
}

@INPROCEEDINGS{VGGSound,
  author={Chen, Honglie and others},
  booktitle={ICASSP},
  title={Vggsound: A Large-Scale Audio-Visual Dataset},
  year={2020},
  number={},
  pages={721-725},
  doi={10.1109/ICASSP40776.2020.9053174}}

@INPROCEEDINGS{aac,
  author={Drossos, Konstantinos and others},
  booktitle={WASPAA}, 
  title={Automated audio captioning with recurrent neural networks}, 
  year={2017},
  number={},
  pages={374-378},
  doi={10.1109/WASPAA.2017.8170058}}

@article{shazeer2020glu,
  title={Glu variants improve transformer},
  author={Shazeer, Noam},
  journal={arXiv preprint arXiv:2002.05202},
  year={2020}
}

@ARTICLE{XACLE2026,
  author={Yuki Okamoto and others},
  title={XACLE Challenge 2026: The first x-to-audio alignment challenge}, 
  journal={UTokyo Repository}, 
  year={2026},
  number={},
  pages={},
}

@article{qwen2.5,
    title   = {Qwen2.5 Technical Report}, 
    author  = {An Yang and others},
    journal = {arXiv preprint arXiv:2412.15115},
    year    = {2024}
}

@article{ziegler2019fine,
  title={Fine-tuning language models from human preferences},
  author={Ziegler, Daniel M and others},
  journal={arXiv preprint arXiv:1909.08593},
  year={2019}
}

@inproceedings{elizalde2023clap,
  title={Clap learning audio concepts from natural language supervision},
  author={Elizalde, Benjamin and others},
  booktitle={ICASSP},
  pages={1--5},
  year={2023},
  organization={IEEE}
}

@article{team2023gemini,
  title={Gemini: a family of highly capable multimodal models},
  author={Team, Gemini and others},
  journal={arXiv preprint arXiv:2312.11805},
  year={2023}
}

@inproceedings{wang2024improving,
  title={Improving text embeddings with large language models},
  author={Wang, Liang and others},
  booktitle={ACL},
  pages={11897--11916},
  year={2024}
}

@article{e5-v,
  title={E5-v: Universal embeddings with multimodal large language models},
  author={Jiang, Ting and others},
  journal={arXiv preprint arXiv:2407.12580},
  year={2024}
}

@inproceedings{BERT,
  title={Bert: Pre-training of deep bidirectional transformers for language understanding},
  author={Devlin, Jacob and others},
  booktitle={NAACL},
  pages={4171--4186},
  year={2019}
}

\end{document}